\documentclass[aps,prb,twocolumn,groupedaddress,showpacs]{revtex4}
\usepackage{bm}
\usepackage{epsf}
\usepackage{amssymb}
\usepackage{amsmath}
\usepackage{graphicx}
\usepackage{rotating}
\usepackage{epsfig}
\usepackage{psfrag}
\usepackage{amsmath}
\usepackage{hyperref}
\usepackage{MnSymbol}
\DeclareMathOperator{\sech}{sech}

\hypersetup{
    bookmarks=true,         
    unicode=false,          
    pdftoolbar=true,        
    pdfmenubar=true,        
    pdffitwindow=true,      
    pdftitle={My title},    
    pdfauthor={Author},     
    pdfsubject={Subject},   
    pdfcreator={Creator},   
    pdfproducer={Producer}, 
    pdfkeywords={keywords}, 
    pdfnewwindow=true,      
    colorlinks=true,       
    linkcolor=red,          
    citecolor=blue,        
    filecolor=magenta,      
    urlcolor=blue           
}

\DeclareMathAlphabet{\bi}{OML}{cmm}{b}{it}

\begin{document}
\def\ea{\textit{et al.}}
\def\bj{\bm{j}}
\def\Im{\mathrm {Im}\;}
\def\Re{\mathrm {Re}\;}
\def\beq{\begin{equation}}
\def\eeq{\end{equation}}
\def\Tr{{\mathrm{Tr}}}
\def\sx{\hat{\sigma}_x}
\def\sy{\hat{\sigma}_y}
\def\sz{\hat{\sigma}_z}
\def\bq{\mathbf{q}}
\def\oc{\omega^{*}}
\def\Hss{\hat{H}_{\mathrm{s}}}
\def\dm{\hat{d}_{k}}
\def\dmd{\hat{d}^{\dagger}_{k}}
\def\ds{\hat{d}_{\sigma}}
\def\dsd{\hat{d}^{\dagger}_{\sigma}}
\def\dsj{\hat{d}_{\sigma}}
\def\dsdj{\hat{d}^{\dagger}_{\sigma}}
\def\tds{\tilde{d}_{\sigma}}
\def\tdsd{\tilde{d}^{\dagger}_{\sigma}}
\def\tdsdj{\tilde{d}^{\dagger}_{\sigma}}
\def\bk{\mathbf{k}}
\def\B{D}
\def\D{d}
\def\Ald{A_{ld}}
\def\Bld{B_{ld}}
\def\Cld{C_{ld}}
\def\P{\hat{\cal{P}}}
\def\Pd{\hat{\cal{P}}^{\dagger}}
\def\bk{\mathbf{k}}
\def\bp{\mathbf{p}}
\def\bq{\mathbf{q}}
\def\dpl{d_{\sigma, \epsilon_0+U}}
\def\d0{d_{\sigma,\epsilon_0}}
\def\dm{d_{\sigma, -\epsilon_0-U}}
\def\dE{d_{\sigma,E}}
\def\dd0{d^{\dagger}_{\sigma,\epsilon_0}}

\title{Thermoelectric performance of strongly-correlated quantum impurity models}
\author{Edward~Taylor}
\affiliation{Chemical Physics Theory Group, Department of Chemistry, University of Toronto,
80 Saint George St. Toronto, Ontario, Canada M5S 3H6}
\author{Dvira~Segal}
\affiliation{Chemical Physics Theory Group, Department of Chemistry, University of Toronto,
80 Saint George St. Toronto, Ontario, Canada M5S 3H6}

\date{\today}

\begin{abstract}
We derive asymptotically exact expressions for the thermopower and figure of merit of a quantum impurity connecting two noninteracting leads in the linear response regime where the chemical potential and temperature differences between the leads are small.  Based on sum rules for the single-particle impurity spectral function, these expressions become exact at high temperatures as well as in the very strongly correlated regime, where the impurity Coulomb repulsion is much larger than the temperature.  Although modest interactions impede thermoelectric performance, a very large Coulomb scale restores the optimal transport properties of noninteracting electrons, albeit renormalized to account for the absence of double occupancy in the impurity.    As with noninteracting electrons, the electronic contribution to the figure of merit is limited only by the spectral broadening that arises from the coupling between the impurity and the leads.  
\end{abstract}
\maketitle

\section{Introduction}
Nearly twenty years ago, Mahan and Sofo elucidated the properties that an ideal thermoelectric material---one 
that can efficiently transform a temperature differential into a voltage---should have~\cite{Mahan96}.  They found that optimal thermoelectric conversion efficiency in the linear response regime, as determined by a large figure of merit, is realized in systems with a transmission function ${\cal{T}}(E)\propto \delta (E-\epsilon_0)$ exhibiting a Dirac-delta function dependence on energy.   In such a system, the electron contribution $\kappa_{el}$ to the thermal conductance at zero charge current vanishes and the figure of merit 
\beq ZT = \frac{G S^2 T}{\kappa_{el} + \kappa_{ph}}\eeq
is only limited by the smallness of $\kappa_{ph}$.   Here $G$ is the electrical conductance, $S$ is the Seebeck coefficient (thermopower), and $\kappa_{el}$ and $\kappa_{ph}$ are the electrical and phonon contributions to the thermal conductance.  

Although realistic systems do not exhibit such transmission, considerable experimental and theoretical work has concentrated on exploring thermoelectric properties of nearly single-level ``impurity'' systems such as quantum dots~\cite{Scheibner05,Costi10,Haupt13,Sothmann15} or molecular junctions~\cite{Paulsson03,Reddy07,Malen10,Yee11, Dubi11,Andergassen11,Lee13,Kim14}.  Such systems can in principle exhibit a transmission that is strongly peaked about this level and hence, promise large figures of merit.  At the same time, these are \emph{small} systems and their coupling to the leads has a significant effect on the transmission function ${\cal{T}}(E)$, yielding an extrinsic broadening $\Gamma$.  Nonetheless, the phonon contribution to the thermal conductance is expected to be small in these systems as compared to that in bulk ones, and there is still great interest in exploring their thermoelectric properties.  (Recent studies have stressed the importance of the phonon contribution $\kappa_{ph}$ in molecules, however~\cite{Stadler11,Burkle15}.)

Beyond numerical studies of the thermoelectric performance of impurity models based on numerical renormalization group~\cite{Costi10,Andergassen11,Costi94} and the nonequilibrium Green's function approach~\cite{Liu10,Ren12}, simple analytic results for thermoelectric coefficients have been found in the ``atomic limit'', in which the broadening $\Gamma\to 0$ is taken to be zero, using e.g., the sequential-tunnelling approximation for transport~\cite{Beenakker92,Koch04,Murphy06}.  In this limit, taking the Coulomb repulsion $U$ to be zero, one recovers the idealized, un-broadened single-level limit of Mahan and Sofo, and correspondingly, an infinite electronic contribution to the figure of merit (i.e., after setting $\kappa_{ph}=0$).  Turning on interactions,  a second satellite level arises due to the interaction energy shift $U$ when two electrons are present in the impurity.  In the atomic limit $\Gamma\to 0$ with strong interactions, $U\gg T$,  Murphy, Mukerjee, and Moore~\cite{Murphy06} showed that the second level becomes unoccupied and hence, irrelevant for transport, meaning that the system again effectively reduces to a single-level one with a diverging electronic figure of merit.  For intermediate coupling strength, $U\sim T$, both levels are active in transport and the figure of merit is not large in general.  

Analogous studies of thermoelectricity in the atomic limit of bulk systems have been undertaken in a pair of well-known papers by Beni~\cite{Beni74} and Chaikin and Beni~\cite{Chaikin76} (see also Ref.~\onlinecite{Mukerjee07}).  As emphasized by Beni~\cite{Beni74}, however, it is nontrivial to perturb away from the atomic limit of transport since transport formally vanishes when $\Gamma = 0$ (for them, the limit where the hopping matrix element between lattice sites vanishes).  For a single-level spin-full impurity, the available Hilbert space for transport formally evolves from having three elements, corresponding to the three possible occupancies, to an infinite number, a large fraction of which must be used in order to have a conserving theory of transport~\cite{Baym61}.  Even when interactions are modest, it is challenging to develop a reliable perturbation expansion of transport quantities in powers of the coupling matrix elements between the impurity and leads (effectively, $\Gamma$).   

In this paper, we use sum rules for the impurity electron spectral function to derive expressions for the thermoelectric coefficients that are asymptotically exact in two regimes for which the broadening is small, but nonzero: First, at high temperatures, greater than $U,\Gamma$ and the bandwidth $\B$ that characterizes the leads.  Second, in the very strongly-correlated, ``narrow-level'' limit where $U\gg T\gg \Gamma$ and $U>\B$.  Our results in this latter regime reproduce the divergent figure of merit in the atomic limit when $U\to \infty$~\cite{Murphy06}.  At the same time, this limit is highly singular and we find
\beq \lim_{\Gamma\to 0} ZT(U\gg T) \sim \frac{(\epsilon_0-\mu)^2}{\Gamma \B},\label{ZTdiv}\eeq
where $\epsilon_0-\mu$ is the difference between the impurity energy and Fermi levels.  This result emphasizes the difficulty of studying transport in this limit and also the crucial role played by the broadening in the thermoelectric performance of impurity models with strong interactions.  

We start in Sec.~\ref{transportsec} by introducing the linear-response formulae for thermoelectric transport coefficients in terms of the single-particle impurity spectral function.   In Sec.~\ref{sumrulesec}, we introduce two sum rules for this spectral function---one that integrates all spectral weight and another which removes the irrelevant weight in the upper Hubbard peak---and show how these can be used to derive results for transport that are asymptotically exact in the regimes elucidated above.  We then calculate explicit forms for these sum rules for the Anderson impurity model in Sec.~\ref{AIMsec}, and use these results to discuss thermoelectric performance in Secs.~\ref{highTsec} and \ref{midTsec}.  Finally, in Sec.~\ref{discussionsec}, we summarize our main results and conclude with a discussion of the implications of our results for quantum dot and molecular junction systems.

\section{Transport coefficients}
\label{transportsec}
Our starting point is the (generalized) Landauer-like expressions for the charge~\cite{Benenti13}
\beq J = \frac{e}{2\pi}\int^{\infty}_{-\infty} dE\; {\cal{T}}(E,T_L,T_R) [f_L(E)-f_R(E)]\label{Jcharge}\eeq
and heat
\beq J_Q = \frac{1}{2\pi} \int^{\infty}_{-\infty} dE\; {\cal{T}}(E,T_L,T_R) (E-\mu_L) [f_L(E)-f_R(E)]\label{Jheat}\eeq
currents through an impurity connecting two leads.  $f_{\nu}(E)\equiv \{\exp[\beta_{\nu}(E-\mu_{\nu})]+1\}^{-1}$ is the Fermi function at the left ($\nu=L$) and right ($\nu=R$) leads, with  temperature $T_{\nu}\equiv \beta^{-1}_{\nu}$ and chemical potential $\mu_{\nu}$.  Unless specified otherwise, throughout this paper we set $\hbar=k_B=1$.  When the coupling between the impurity and the left and right leads are proportional, Meir and Wingreen showed that (\ref{Jcharge}) is an exact expression with the identification of the generalized transmission function with the spectral function $A(E)$ of the impurity electrons~\cite{MW}:
\beq {\cal{T}}(E,T_L,T_R) = (\pi \gamma/2)\Gamma(E)A(E-\mu),\label{TMW}\eeq
where 
\beq \Gamma \equiv \Gamma_L + \Gamma_R\eeq
is the sum of the broadenings [assumed in (\ref{TMW}) to be purely real] at the left and right leads,
\beq \gamma \equiv 4\Gamma_L\Gamma_R/\Gamma^2 \eeq
is an asymmetry parameter that deviates from unity when $\Gamma_L\neq \Gamma_R$, and
\beq A(E) \equiv -\frac{1}{\pi} \sum_{\sigma}\mathrm{Im}G^{\mathrm{ret}}_{\sigma}(E)\eeq
is the spectral function for both spin species $\sigma$ of impurity electrons with retarded Green's function $G^{\mathrm{ret}}(E)$.  It trivially follows from the analysis in Ref.~\onlinecite{MW} that (\ref{Jheat}) is also exact with the identification in (\ref{TMW})~\cite{Andergassen11,Kim02}.   

The precise form of the spectral function $A(E)$ is specified by the details of the coupling between the impurity and the leads.  To be specific, suppose
\beq \hat{V} =\sum_{\bk\nu\sigma}\left[V_{\bk \nu}\hat{c}^{\dagger}_{\bk\nu\sigma}\hat{d}_{\sigma} + \mathrm{h.c.}\right]\label{V}\eeq
couples the single-level impurity electrons characterized by $\hat{d}_{\sigma}$ to the non-interacting $\nu=L,R$ leads,   
and
\beq\hat{H}_l = \sum_{\bk\nu\sigma}(\epsilon_{\bk\nu}-\mu_{\nu})\hat{n}_{\bk\nu\sigma}\label{Hc}\eeq 
describes noninteracting electrons with momentum distribution $\hat{n}_{\bk\nu\sigma}\equiv \hat{c}^{\dagger}_{\bk\nu\sigma}\hat{c}_{\bk\nu\sigma}$ in these leads. For this model,
$A(E)$ is given by
\beq A(E) = \sum_{\sigma}\sum_{a,b}(P_a+P_b)\langle b|\hat{d}^{\dagger}_{\sigma}|a\rangle\langle a|\hat{d}_{\sigma}|b\rangle \delta(E-E_b+E_a),\label{A}\eeq 
and the broadening is
\beq \Gamma_{\nu}(E) = 2\pi \sum_{\bk}\delta(E-\epsilon_{\bk \nu})|V_{\bk \nu}|^2.\label{Gamma}\eeq  
The states $|a\rangle, |b\rangle$ in the spectral representation (\ref{A}) are the exact eigenstates of the many-body grand-canonical hamiltonian, $\hat{H}|a\rangle = E_a|a\rangle$.  Within linear response $\mu_L\simeq \mu_R$ and $T_L\simeq T_R$, $P_a \equiv \exp(-\beta E_a)/{\cal{Z}}$, with ${\cal{Z}}$ the grand canonical partition function.   $\hat{H}$ includes (\ref{V}) and (\ref{Hc}) as well as the as-yet-unspecified impurity hamiltonian.   We will restrict ourselves in what follows to the linear response regime.

Equations (\ref{Jcharge}), (\ref{Jheat}), and (\ref{TMW}) lead to the following expressions for the transport coefficients~\cite{Mahan96,Benenti13} in the linear response regime for an energy-independent broadening $\Gamma_{\nu}(E)=\Gamma_{\nu}$ (momentarily restoring $\hbar$ and $k_B$):
\beq G = \frac{e^2\gamma\Gamma}{16\hbar k_B T}M_0,\label{Ge}\eeq
\beq \kappa_{el} = \frac{\gamma\Gamma}{16\hbar k_B T^2} \left(M_2-\frac{M^2_1}{M_0}\right),\label{GQ}\eeq
\beq S = \frac{1}{eT}\frac{M_1}{M_0},\label{S}\eeq
and
\beq ZT = \frac{M^2_1}{M_0(M_2-M^2_1/M_0)},\label{ZT}\eeq  
defined in terms of the integral expression
\beq M_n \equiv \int^{\infty}_{-\infty} dE E^n \sech^2(\beta E/2) A(E).\label{M}\eeq
In arriving at these expressions, we have shifted $E\to E+\mu$ in the linear-response limit $\mu_L=\mu_R=\mu$ of the expressions (\ref{Jcharge}) and (\ref{Jheat}) for the currents, meaning that the spectral function $A(E)$ is the one defined in (\ref{A}).  An energy-independent broadening is appropriate for leads that exhibit good metallic behaviour with a broad bandwidth, generally much larger than $\Gamma$.  Had we not made this assumption, the broadening $\Gamma(E)$ would have entered the integral expression (\ref{M}).  (As we explain below, this situation can also be dealt with using the methods developed in this paper, although the resulting calculations would be more complicated.) The above expressions will form the basis of our analysis in the remainder of this paper.

Before closing this section, we briefly comment on the major result of Mahan and Sofo who, as noted earlier, showed that the electronic contribution to the figure of merit would diverge in a system described by a single, un-broadened energy level.  In the language of this section, this corresponds to having a Dirac-delta function spectral function
\beq A(E)\propto \delta(E-\epsilon_0).\label{MSA}\eeq
Using this in the above, the thermopower is (restoring $k_B$) $S = (k_B/e)(\epsilon_0/T)$.  Crucially, using (\ref{MSA}) in (\ref{M}), one finds
\beq M_2 = \frac{M^2_1}{M_0},\label{divergingkappacond}\eeq
and the electronic contribution to the thermal conductance vanishes identically at \emph{all temperatures}, with the result that the figure of merit is divergent at all temperatures.  

Moving away from the limit (\ref{MSA}) of having a single un-broadened energy level, a small thermal conductance can more generally be understood as expressing a small variance in the single-particle energy.  Defining 
\beq \llangle \cdots\rrangle_F\equiv \frac{\int dE (\cdots)A(E)F(E)}{\int dE A(E)F(E)},\label{avg}\eeq we see that \beq M_2-\frac{M^2_1}{M_0} = M_0[\llangle E^2\rrangle_F-\llangle E\rrangle^2_{F}],\label{variance}\eeq 
with $F=\sech^2(\beta E/2)$.  Hence, the ratio
\beq \frac{\kappa_{el}}{GT} = \frac{1}{e^2 T^2}\left(\llangle E^2\rrangle_F-\llangle E\rrangle^2_{F}\right)_{F=\sech^2(\beta E/2)},\label{conductivityratio}\eeq
which defines the Lorenz number in bulk systems, provides a direct measure of the variance in the single-particle energies clustered within $T$ of the Fermi level (because of the form of $F$).  It is thus clear why a system with a single un-broadened energy level would be optimal, having zero variance in its energy levels; see also Eq.~(18) in Mahan and Sofo~\cite{Mahan96}.  This result also presages our main conclusion: Even though an extremely large $U$ can eliminate the energy-level variance due to interactions, unless the broadening $\Gamma$ due to coupling with the leads can be made smaller than the temperature (not currently the situation in experiments~\cite{Paulsson03,Quek07}), the energy variance will be greater that $T^2$, and one should only expect modest thermoelectric performance.

\section{Sum rule expressions for transport coefficients}
\label{sumrulesec}
The evaluation of the spectral function $A(E)$ that enters the expressions for the thermoelectric coefficients is a challenging many-body problem.  In this paper, we propose an alternative, evaluating instead
\beq M^{T}_n \equiv \int^{\infty}_{-\infty} dE E^n A(E)\label{Minf}\eeq
and
\beq \tilde{M}_n(\Omega_c) \equiv \int^{\Omega_c}_{-\infty} dE E^n A(E).\label{Mc}\eeq
The first of these is a well-known \emph{sum rule}, and can straightforwardly be evaluated in terms of commutators involving electron creation and annihilation operators and the hamiltonian (see e.g., Refs.~\onlinecite{Harris67,White91}).  Integrating the product of $E^n$ and (\ref{A}) leads to (see Appendix~\ref{spectralsec})
\beq M^{T}_n =  \sum_{\sigma}\langle \{\hat{d}^{\dagger}_{\sigma},[\hat{d}_{\sigma},\hat{H}]_n\}\rangle. \label{Minfexp}\eeq
Here $[\hat{d}_{\sigma},\hat{H}]_n$ is a nested commutator with $[\hat{d}_{\sigma},\hat{H}]_0= \hat{d}_{\sigma}$ the zeroth-order commutator, $[\hat{d}_{\sigma},\hat{H}]_1 = [\hat{d}_{\sigma},\hat{H}]$ the first-order commutator, $[\hat{d}_{\sigma},\hat{H}]_2 = [[\hat{d}_{\sigma},\hat{H}],\hat{H}]$, and so on.  

The second expression, (\ref{Mc}), involving an energy cutoff $\Omega_c$, is evaluated by projecting out the states with energy above $\Omega_c$.  Taking $\Gamma\ll \Omega_c\ll U$, this amounts to deriving an effective operator $\d0$ for the lower Hubbard peak.  The resulting sum rule for the spectral weight contained in the lower Hubbard peak of a single-level impurity model is 
\beq \tilde{M}_n =  \sum_{\sigma}\langle \{\dd0,[\d0,\hat{H}]_n\}\rangle. \label{Mtilde}\eeq
The precise choice of operator $\d0$ depends on the model under consideration; its derivation for the Anderson impurity model coupled to two leads will be given in the next section.  We could equally derive a sum rule for the spectral weight contained in the upper Hubbard peak, as would be relevant for the situation when the chemical potential is close to this peak.  In what follows, we will assume that the Fermi level is close to the lower Hubbard peak, however.

\begin{figure}
\includegraphics[width=0.7\linewidth]{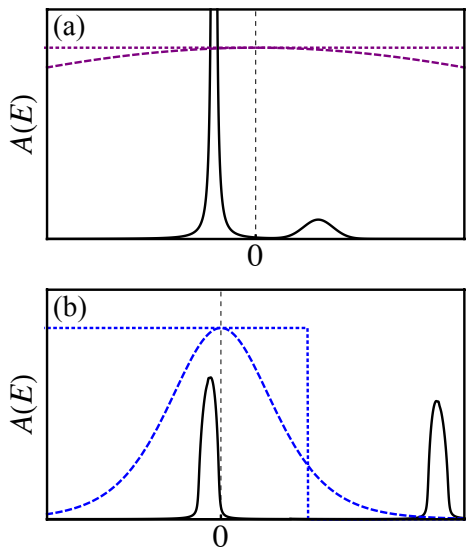}
\caption{Schematic plot of the spectral function $A(E)$ at high temperatures $T\gg U,\Gamma$ (a) and for strong correlations $\Gamma \lesssim T\ll U$ (b) relative to the lead Fermi level (denoted here by $E=0$).  Also shown by the dashed lines is the thermal weight function $\sech^2(\beta E/2)$ that enters the moments $M_n$; the dotted lines show the weight functions---unity and the step function $\Theta(\Omega_c-E)$---involved with the sum rules.}
\label{spectra}
\end{figure}

The sum rule (\ref{Minf}) provides a rigorous upper bound on $M_n$ when $n$ is even: 
\beq M_n\leq M^{T}_n \;\;n=0,2,...\;\;\;\forall T,\label{momentbound}\eeq
since $A(E)\geq 0$ $\forall E$ and $\sech^2(x)\leq 1\; \forall x$.  This immediately allows us to write down upper bounds on the charge conductance
\beq G \leq \frac{e^2\Gamma}{8\hbar T}M^{T}_0 \eeq
and, making use of the fact that $M^2_1/M_0\geq 0$~\cite{kappacomment}, thermal conductance $\kappa_{el}$ at zero charge current:
\beq \kappa_{el} \leq \frac{\Gamma}{2\hbar  T^2}M^{T}_2.\eeq
Related bounds for the thermal conductance due to bosonic excitations (e.g., phonons) were derived in Ref.~\onlinecite{Taylor15}.  

Because they involve a ratio of moments, (\ref{momentbound}) does not obviously translate to bounds for the Seebeck coefficient $S$ and figure of merit $ZT$.  The usefulness of the sum rules is that they provide asymptotically exact approximations to the transport coefficients at high-temperatures, including the asymptotic limit $T\gg U,\Gamma,\B$, and also in the strongly-correlated regime $\Gamma \lesssim T\ll U$.  
In Fig.~\ref{spectra}, we show two schematic plots of the impurity spectral function along with the weight function $\sech^2(\beta E/2)$ that enters the exact transport moments $M_n$.  In Fig.~\ref{spectra}(a), the temperature is much larger than the Coulomb repulsion $U$ as well as $\Gamma$; the majority of the spectral weight arises in a Lorentzian centred around the Fermi level of width $\sim \Gamma$.  A small peak---the weight of which vanishes with decreasing $U/T$---arises at $U$ corresponding to the energy of a doubly occupied impurity.   In this regime, the $\sech^2(\beta E/2)$ function is essentially unity wherever there is nonzero spectral weight and one can replace $M_n$ by the high-temperature asymptotes $M^{T}_n$, with the result that the thermoelectric transport coefficients calculated using these sum rules become asymptotically exact:
\beq \lim_{T\gg U,\Gamma,\B}\left\{G,\kappa_{el},S,ZT\right\}[M_n] = \left\{G,\kappa_{el},S,ZT\right\}[M^{T}_n].\label{HT}\eeq
We formally require that $T\ll \B$ be much larger than the lead bandwidth since the spectral function scales as $A(E)\propto E^{-2}$ for large $E$, meaning that $M^{T}_1$ and $M^{T}_2$ cannot well-approximate $M_1$ and $M_2$ unless $\B\lesssim T$.  

Fig.~\ref{spectra}(b) shows the spectral function in the strongly-correlated regime $\Gamma \ll T\ll U$.  Here two Hubbard peaks develop, the lower one centred at some single-particle energy scale, $\epsilon_0$, and the upper one centred at $\epsilon_0+U$.  When the Fermi level (denoted by the zero in these plots) is within $\sim T$ of e.g. the lower Hubbard peak, the $\sech^2(\beta E/2)$ function excludes the upper Hubbard peak.  Because there is negligible spectral weight outside the two peaks when $U\gg \Gamma$~\cite{comment}, in this regime $M_n$ can be replaced by $\tilde{M}_n(\Omega_c)$, where $\Gamma \ll \Omega_c\ll U$ and the transport coefficients again admit asymptotically exact sum rule values:
\begin{align} \lim_{\Gamma \ll T\ll U}&\left\{G,\kappa_{el},S,ZT\right\}[M_n] =\nonumber\\& \left\{G,\kappa_{el},S,ZT\right\}[\tilde{M}_n(\Gamma\ll \Omega_c\ll U)].\label{HM}\end{align}
Even though (\ref{HM}) is asymptotically exact in the limit $\Gamma \ll T$, it is evident that these results will remain qualitatively valid for $\Gamma \lesssim T$.  

As noted earlier, we could also consider the thermoelectric performance when the Fermi level is close to the upper Hubbard peak by considering the sum rule for this peak.  Apart from effects that break particle-hole  symmetry (enhanced by molecular vibrational modes~\cite{Ren12}), the thermoelectric performance will be the same in both cases.  

\begin{figure}
\includegraphics[width=0.85\linewidth]{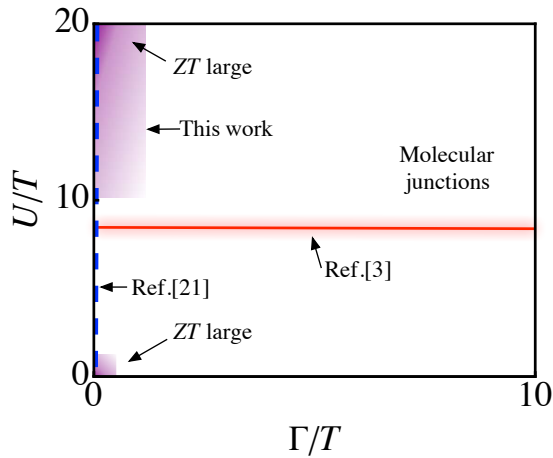}
\caption{Plot of the regimes where our results hold ($U\gg T\gtrsim \Gamma$ and $T\gg U,\Gamma,\B$), in comparison to the regimes studied in the sequential tunnelling approximation of Ref.~\onlinecite{Murphy06} (all $U$, $\Gamma/T\to 0$) and the numerical renormalization group calculations of Ref.~\onlinecite{Costi10} ($U/\Gamma = 8$, all $\Gamma$).  Note that the latter calculation could have been extended to any region shown above.  Also indicated are the regimes ($\Gamma, U \ll T$ and $\Gamma\ll T, U\gg T$) where the electronic contribution to the figure of merit becomes large as well as the parameter regime of typical molecular junction experiments.  Although the value of $U/T$ is unknown, $\Gamma$ is typically $\gtrsim 10T$~\cite{Quek07,Paulsson03}.}
\label{regimes}
\end{figure}

In Figure~\ref{regimes}, we plot the regimes in the parameter space spanned by $U/T,\Gamma/T$ where our results hold.  Also shown is the regime studied by Murphy, Mukerjee, and Moore~\cite{Murphy06}, as well as that studied using NRG by Costi and Zlati\'c~\cite{Costi10}, although NRG can study the entire region shown in this figure.  Even though it might seem that we are dealing with a highly restricted region of parameter space---and one that is moreover far from most current experiments involving molecular junctions for which $\Gamma\sim {\cal{O}}(10 T)$~\cite{Quek07,Paulsson03}---we re\"emphasize that it is \emph{only when $\Gamma\ll T$ that a substantial figure of merit can be realized. }  We also note that this regime is well away from the Kondo regime $T\lesssim T_K$, since $T_K\ll \Gamma$~\cite{Costi10,Hewsonbook}

In the remainder of this paper, we investigate the implications of (\ref{HT}) and (\ref{HM}) for an archetypal model of strongly correlated ``impurity'' electrons, the Anderson impurity model, generalized to include the coupling between an impurity and two leads.  These sum-rule results have been derived assuming a constant broadening function, as is almost always taken to be the case.  One can also consider energy-dependent broadenings: for simple power-law forms such as $\Gamma(E) = \Gamma E^l$ with $l$ and integer, the resulting transport coefficients simply involve higher-moment sum rules ($n\to n+l$).  For more complicated dependencies, one must resort to the operator-product expansion technique~\cite{OPE} of Wilson, Kadanoff, and Polyakov.

\section{Sum rules for the Anderson impurity model} 
\label{AIMsec}
The single-level Anderson impurity model~\cite{Hewsonbook} (AIM) for an impurity coupled to two leads is 
\beq\hat{H} = \hat{H}_d + \hat{H}_l+\hat{V}.\label{AIM}\eeq 
Here
\beq \hat{H}_d = (\epsilon_0-\mu)\sum_{\sigma}\hat{d}^{\dagger}_{\sigma}\hat{d}_{\sigma} + U\hat{n}_{\uparrow}\hat{n}_{\downarrow}\label{Hd}\eeq
is the impurity hamiltonian and  $H_l$ and $\hat{V}$ are given by  (\ref{Hc}) and (\ref{V}), the former with $\mu_{\nu}=\mu$, as appropriate in the linear response regime.

\subsection{Sum rules for the entire spectrum}
The results of a straightforward calculation using (\ref{Minfexp}) and (\ref{AIM}) are
\beq M^{T}_0 =2,\label{Minf0}\eeq
\beq M^{T}_1 = 2(\epsilon_0-\mu + Un_d/2),\label{Minf1}\eeq
and 
\beq	 M^{T}_2= 2(\epsilon_0-\mu)^2+2(\epsilon_0-\mu)U n_d + U^2 n_d +2\sum_{\bk\nu}|V_{\bk\nu}|^2.\label{Minf2}\eeq
Here $n_d \equiv \sum_{\sigma}n_{\sigma}$ is the total occupancy of the impurity.  The factor of two in (\ref{Minf0}) (and elsewhere in the above) counts the number of single-particle states: two for the single-level AIM with two spins.  Taking the $V_{\bk\nu}$ to be independent of momentum, 
\begin{align} \sum_{\bk\nu}|V_{\bk\nu}|^2 &= \sum_{\nu}|V_{\nu}|^2 N_0\int^{\B}_{-\B} dE\nonumber\\&=2\B N_0\sum_{\nu}|V_{\nu}|^2.\label{bandwidthlimapprox} \end{align}
Here we have taken the lead DOS $N_{\sigma}(E)=N_0$ per spin to be constant (and independent of spin) with $\B$ the half-bandwidth.  Making the same assumptions to evaluate the broadening (\ref{AIM}), it becomes
\beq \Gamma_{\nu} = 2\pi N_0 |V_{\nu}|^2,\label{GammaAIM2}\eeq
and thus
\beq  \sum_{\bk\nu}|V_{\bk\nu}|^2 =\B\Gamma/\pi.\label{Bld0}\eeq

As noted in (\ref{variance}), the combination $M^{T}_2-(M^{T}_1)^2/M^{T}_0 = M^{T}_0[\llangle E^2\rrangle_{F=1}-\llangle E\rrangle^{2}_{F=1}]$ provides a measure of the variance in the single-particle energies, in this case over the entirety of the spectral weight since there is no thermal weighting function $\sech^2(\beta E/2)$.  In order to have a small electronic contribution to the thermal conductance and hence, a substantial figure of merit, the energy variance in the system needs to be small.  Equations (\ref{Minf0})-(\ref{Minf2}) give
\beq M^{T}_2-\frac{(M^{T}_1)^2}{M^{T}_0} = \frac{U^2}{2}n_d(2-n_d)+2\sum_{\bk\nu}|V_{\bk\nu}|^2.\label{highTvariance}\eeq
The first term on the right-hand side gives the variance in the interaction energy per particle (proportional to $U^2$ times the compressibility), while the second term describes the variance in the kinetic energy per particle.  One can understood this result in terms of Fig.~~\ref{spectra}: with two peaks separated by $U$, the energy variance is clearly $\sim U^2$.  The variance in a \emph{single band} (either one) however, is $\sim\sum_{\bk\nu}|V_{\bk\nu}|^2\sim \Gamma \B$.   

At temperatures greater than both $\Gamma$ and $U$, the variance is of course large, and (\ref{Minf2}) indicates a correspondingly large $\kappa_{el}$.  On the other hand, for lower temperatures, $T\ll U$, assuming the Fermi level is close to the lower Hubbard peak, the $\sech^2(\beta E)$ factor in $M_n$ is sensitive only to the spectral weight in that peak.  Correspondingly, $\kappa_{el}$ is only sensitive to the variance $\sim\Gamma\B$ in this peak.  This notion can be made precise by projecting out the spectral weight contained in the upper Hubbard peak from the sum rules, as in (\ref{Mc}).  We turn to this now.

\subsection{Projected sum rules for the lower Hubbard peak}
Relevant to the case shown in Fig.~\ref{spectra}(b), we now calculate the sum rule for the lower Hubbard peak (LHP), obtained by projecting out the doubly-occupied states that give rise to the upper Hubbard peak (UHP).  Our approach here follows closely that used in studies of the bulk electron Hubbard model~\cite{Harris67,Eskes94,Randeria05} and also the AIM itself by Schrieffer and Wolff~\cite{Schrieffer66}.  To remove the spectral weight associated with the UHP, we need a spectral decomposition for the impurity electron operator of the form
\beq d_{\sigma} = \d0 + \dpl.\label{spectraldecomp}\eeq
Here $\dE$ describes an annihilation process that lowers the impurity energy by $\sim E$ modulo $\Gamma$, i.e., $\langle b|\dE |a\rangle = 0$ unless $E_a-E_b = E+{\cal{O}}(\Gamma)$.  Once a suitable expression is found for $\d0$, one can proceed to calculate the projected sum rules given by (\ref{Mtilde}).

When $\Gamma=0$, the decomposition is particularly simple since the occupancies $\hat{n}_{\sigma}$ are good quantum numbers, and, defining $\bar{\sigma} = -\sigma$, 
\beq \d0 = (1-\hat{n}_{\bar{\sigma}})\hat{d}_{\sigma}\label{d0}\eeq
and
\beq \dpl = \hat{n}_{\bar{\sigma}}\hat{d}_{\sigma}.\label{dpl}\eeq
When $\Gamma \neq 0$, we seek a \emph{renormalized} operator $\tilde{d}_{\sigma}$ such that  $\tilde{n}_{\sigma}$ is a good quantum number of the bare hamiltonian $\hat{H}$, and $[\tilde{n}_{\sigma},\hat{H}] = 0$.  
This allows us to undertake a similar spectral decomposition as (\ref{spectraldecomp}) and identify the renormalized operator $\d0$ for the LHP needed for the sum rules.     

Since $\tilde{n}_{\sigma}$ commutes with $\tilde{H}_0 \equiv \hat{H}_0(\hat{d}_{\sigma}\to \tilde{d}_{\sigma}, \hat{c}_{\bk\nu\sigma}\to \tilde{c}_{\bk\nu\sigma})$, where
\beq \hat{H}_0 \equiv \hat{H}_d - \mu \sum_{\bk\nu}\hat{n}_{\bk\nu\sigma}\label{H0}\eeq 
is the total hamiltonian less all kinetic energies, $[\tilde{n}_{\sigma},\hat{H}] = 0$ implies $[\tilde{H}_0,\hat{H}]=0$.  
Relating the renormalized operators to the original operators by a unitary transformation (not to be confused with the thermopower $S$, but in keeping with widely-used notation)~\cite{Schrieffer66,Harris67,MacDonald88,Eskes94,Randeria05} $\hat{d}_{\sigma}   = e^{S}\tilde{d}_{\sigma}e^{-S}$ and $\hat{c}_{\bk\nu\sigma} = e^S\tilde{c}_{\bk\nu\sigma}e^{-S}$, this last condition amounts to finding a transformation operator $S$ that satisfies \beq [\hat{H},\tilde{H}_0]=0\Rightarrow [e^{S}\tilde{H}e^{-S},\tilde{H}_0]=0.\label{cond1}\eeq
Here, and in the remaining, we use a tilde to indicate a transformed operator.

The fact that $\tilde{n}_{\sigma}$ is a good quantum number means that---analogous to (\ref{spectraldecomp})---we can decompose any operator $\tilde{O} = \tilde{O}(\tilde{d}_{\sigma},\tilde{c}_{\bk\nu\sigma})$ in terms of the eigenstates $\tilde{n}_{d} = \sum_{\sigma}\tilde{n}_{\sigma} = 0,1,2$, or equivalently, the corresponding energy changes $E = \pm \epsilon_0, \pm (\epsilon_0+U)$ that they effect, modulo corrections ${\cal{O}}(\Gamma)$:
\beq \tilde{O} = \tilde{O}_{\pm \epsilon_0}+\tilde{O}_{\pm(\epsilon_0+U)}.\label{spectraldecomp2}\eeq
$\tilde{O}_{\pm E}$ operators are related by hermitian conjugacy:
\beq \tilde{O}_{-E} = \tilde{O}^{\dagger}_{E}.\label{conjcond}\eeq
Equation~(\ref{spectraldecomp}) is a special example of this more general spectral decomposition.    
Because the eigenvalues $\tilde{n}_d$ of $\tilde{H}_0$ are good quantum numbers, so are the energy changes $E$ defined above, with the result that 
\beq [\tilde{O}_{E}, \tilde{H}_0] = E\tilde{O}_E+{\cal{O}}(\Gamma),\;\;E = \pm\epsilon_0,\pm (\epsilon_0+U).\label{useful}\eeq

Equation~(\ref{useful}) allows us to determine $S$ and $\d0$ to a specified order in $V/(\epsilon_0+U)$, where $V$ is the characteristic size of the in-general momentum dependent coupling $V_{\bk\nu}$.  
We expand $\hat{H}= \exp(S)\tilde{H}\exp(-S)$ as $\hat{H} = \tilde{H} + [S,\tilde{H}] + \cdots$ and $S = S^{(1)}+S^{(2)}+\cdots$ in powers of $V/(\epsilon_0+U)$ and use this in (\ref{cond1}).  Requiring that $S$ eliminates the $\tilde{H}_{\pm(\epsilon_0+U)}$ terms in the spectral decomposition (\ref{spectraldecomp2}) of $\tilde{H}$ (thereby removing the associated spectral weight) and using (\ref{useful}), (\ref{cond1}) reduces to
\beq [S^{(1)},\tilde{H}_d] = -\tilde{T}_{\epsilon_0+U}-\tilde{T}_{-\epsilon_0-U}.\label{Scond}\eeq
Here $\tilde{T}_E$ are the terms in the spectral decomposition of the kinetic energy contribution $\tilde{T}\equiv \tilde{H}-\tilde{H}_0$ to the hamiltonian; note that these are the only terms in the hamiltonian that can change the impurity energy by $\sim \pm(\epsilon_0+U)$: $\tilde{H}_{\pm(\epsilon_0+U)} = \tilde{T}_{\pm(\epsilon_0+U)}$.  Explicitly,  
\beq \tilde{T}_{\epsilon_0} = \sum_{\bk\nu\sigma}\epsilon_{\bk\nu}\tilde{n}_{\bk\nu\sigma} + \sum_{\bk\nu\sigma}\left[V_{\bk\nu}\tilde{c}^{\dagger}_{\bk\nu\sigma}\tilde{d}_{\sigma}(1-\tilde{n}_{\bar{\sigma}}) + \mathrm{H.c.}\right]\label{T0}\eeq
and
\beq \tilde{T}_{\epsilon_0+U} = \sum_{\bk\nu\sigma}V^{*}_{\bk\nu}\tilde{n}_{\bar{\sigma}}\tilde{d}^{\dagger}_{\sigma}\tilde{c}_{\bk\nu\sigma}.\label{TU}\eeq

Using (\ref{TU}) and (\ref{conjcond}) in (\ref{Scond}), one finds
\beq S^{(1)} = -\frac{1}{\epsilon_0+U}\left[\tilde{T}_{-\epsilon_0-U}-\tilde{T}_{\epsilon_0+U}\right].\label{S1}\eeq
This is essentially~\cite{SWcomment} just the Schrieffer--Wolff transformation used by Schrieffer and Wolff to derive the Kondo hamiltonian as the effective low-energy description of the AIM\cite{Schrieffer66}.  Apart from the presence of the additional single-particle energy scale $\epsilon_0$ in this expression, it is also the same result as obtained for the bulk Hubbard model, with $\tilde{V}$ replaced by inter-site tunnelling; see e.g., Eq.~(7) in Ref.~\onlinecite{MacDonald88} and Eq.~(10) in Ref.~\onlinecite{Eskes94}.  

Having determined the form of $S^{(1)}$, we can now obtain an expression for the renormalized LHP operator appearing in (\ref{spectraldecomp}).  By expanding $\hat{d}_{\sigma} = \exp(S)\tilde{d}_{\sigma}\exp(-S) = \tilde{d}_{\sigma} + [S^{(1)},\tilde{d}_{\sigma}]+\cdots$ and taking the spectral decomposition of both sides of the resulting expression, one finds perturbative expressions for the spectral decomposition $d_{\sigma,\epsilon_0}$ and $d_{\sigma,\epsilon_0+U}$ of the \emph{original} operator; c.f. (\ref{spectraldecomp}).  
Using (\ref{Scond}) and equating terms of the same order in $E$ gives
\begin{align} \d0 =(1-\tilde{n}_{\bar{\sigma}})\tilde{d}_{\sigma} +& \frac{1}{\epsilon_0+U}\left[\tilde{T}_{\epsilon_0+U},\tilde{d}_{\sigma,-\epsilon_0-U}\right]\nonumber\\ &- \frac{1}{\epsilon_0+U}\left[\tilde{T}_{-\epsilon_0-U},\tilde{d}_{\sigma,-\epsilon_0-U}\right],\label{d02}\end{align}
for the LHP operator.  Explicit evaluation of this expression results in
\begin{align} \d0 =& (1-\tilde{n}_{\bar{\sigma}})\tilde{d}_{\sigma} - \frac{1}{\epsilon_0+U}(\tilde{n}_{\sigma}+\tilde{n}_{\bar{\sigma}})\tilde{d}_{\sigma}\sum_{\bk\nu}V^{*}_{\bk\nu}\tilde{d}^{\dagger}_{\bar{\sigma}}\tilde{c}_{\bk\nu\bar{\sigma}}\nonumber\\
&-\frac{1}{\epsilon_0+U}\sum_{\bk\nu}V^{*}_{\bk\nu}\tilde{n}_{\bar{\sigma}}\tilde{c}_{\bk\nu\sigma} + {\cal{O}}[V/(\epsilon_0+U)]^2.\label{d03}\end{align}
In the limit $U\to \infty$, (\ref{d03}) reduces to (\ref{d0}).  

Using (\ref{d03}) to evaluate the first few sum rules (\ref{Mtilde}), making use of the fact that there are no doubly-occupied states in the transformed basis, $\tilde{d}^{\dagger}_{\sigma}\tilde{d}^{\dagger}_{\bar{\sigma}} = 0$, one finds after some straightforward but laborious algebra
\beq \tilde{M}_0 = 2-n_d - \frac{1}{\epsilon_0+U}\sum_{\bk\nu\sigma}\left[V^{*}_{\bk\nu}\langle \tilde{d}^{\dagger}_{\sigma}\tilde{c}_{\bk\nu\sigma}\rangle + \mathrm{H.c.}\right] + {\cal{O}}\left(\tfrac{V}{\epsilon_0+U}\right)^2,\label{M0t}\eeq
\beq \tilde{M}_1 = (\epsilon_0-\mu)\tilde{M}_0 + \sum_{\bk\nu\sigma}V^{*}_{\bk\nu}\langle \tilde{d}^{\dagger}_{\sigma}\tilde{c}_{\bk\nu\sigma}\rangle + {\cal{O}}\left(\tfrac{V^2}{\epsilon_0+U}\right),\label{M1t}\eeq
and
\begin{align} &\tilde{M}_2 =(\epsilon_0-\mu)\tilde{M}_1 + (2-n_d)\sum_{\bk\nu}|V_{\bk\nu}|^2\nonumber\\& - \sum_{\bk\nu\sigma}V^{*}_{\bk\nu}(\epsilon_{\bk\nu}-\mu)\langle \tilde{d}^{\dagger}_{\sigma}\tilde{c}_{\bk\nu\sigma}\rangle\nonumber\\&  +\left\langle \left(\sum_{\bk\nu\sigma}V^{*}_{\bk\nu}\tilde{d}^{\dagger}_{\sigma}\tilde{c}_{\bk\nu\sigma}\right)\left(\sum_{\bk'\nu'\sigma'}V_{\bk'\nu'}\tilde{c}^{\dagger}_{\bk'\nu'\sigma'}\tilde{d}_{\sigma'}\right)\right\rangle + {\cal{O}}\left(\tfrac{V^3}{\epsilon_0+U}\right). \label{M2t}\end{align}
The analogue of (\ref{M0t}) for the bulk Hubbard model has been derived in Refs.~\onlinecite{Harris67,Eskes94,Randeria05}.  Note that e.g.
\beq \sum_{\bk\nu\sigma}V^{*}_{\bk\nu}\langle \tilde{d}^{\dagger}_{\sigma}\tilde{c}_{\bk\nu\sigma}\rangle = \sum_{\bk\nu\sigma}V_{\bk\nu}\langle \tilde{c}^{\dagger}_{\bk\nu\sigma}\tilde{d}_{\sigma}\rangle\label{reality}\eeq
is purely real in equilibrium, as required to have no charge or heat current.  Defining
\beq \hat{{\cal{V}}} \equiv \sum_{\bk\nu\sigma}V_{\bk\nu}\hat{c}^{\dagger}_{\bk\nu\sigma}\hat{d}_{\sigma},\label{v}\eeq
such that $\hat{V} \equiv \hat{\cal{V}} + \hat{\cal{V}}^{\dagger}$ is the lead-impurity coupling (\ref{V}), these give the following result for the variance in the LHP:
\begin{align} \tilde{M}_2 - \frac{\tilde{M}^2_1}{\tilde{M}_0} = &(2-n_d)\sum_{\bk\nu}|V_{\bk\nu}|^2 -\sum_{\bk\nu\sigma}V^{*}_{\bk\nu}(\epsilon_{\bk\nu}-\mu)\langle \tilde{d}^{\dagger}_{\sigma}\tilde{c}_{\bk\nu\sigma}\rangle\nonumber\\ &+ \langle \tilde{{\cal{V}}}^2\rangle - \frac{1}{(2-n_d)}\langle \tilde{{\cal{V}}}\rangle^2.\label{variancet}\end{align}

Comparing the full sum rules (\ref{Minf0})-(\ref{Minf2}) with the projected ones (\ref{M0t})-(\ref{M2t}) and also the variances, (\ref{variance}) and (\ref{variancet}), one sees that the projection has two primary effects:  First, it removes spectral weight associated with the UHP and the large Coulomb repulsion $U$.  Second, it renormalizes the spectral weight from $2$ (the total spectral weight) to $\sim 2-n_d$, the low-energy spectral weight in the LHP.   The variance (\ref{variancet}) is essentially the variance of a \emph{non-interacting single level}, broadened by $\sum_{\bk\nu}|V_{\bk\nu}|^2$ and $\langle \tilde{\cal{V}}\rangle$; the effects of interactions only enter through the spectral weight renormalization $2\to 2-n_d$ and their effect on the coherence $\langle\tilde{d}^{\dagger}_{\sigma}\tilde{c}_{\bk\nu\sigma}\rangle$.  To the orders in $V/(\epsilon_0+U)$ shown, one can replace the transformed operators $\tilde{d}_{\sigma}$, $\tilde{c}_{\bk\nu\sigma}$ in the above with the original operators, $\hat{d}_{\sigma}$, $\hat{c}_{\bk\nu\sigma}$, mindful, however, of the restriction that there be no doubly-occupied states.  

We now use the above results to give insight into the thermoelectric performance of strongly correlated materials in the high-temperature and strongly-correlated regimes.  

\section{Thermoelectric performance at high temperatures $T\gg U$}
\label{highTsec}
Using the results from the previous section, we now evaluate the exact high-temperature limiting values of the thermopower $S$ and figure of merit $ZT$ for the Anderson impurity model.
In the high temperature limit, $n_d(T\gg U,\Gamma) \to 1$.  Using this in, (\ref{Minf0})-(\ref{Minf2}), the high-temperature values for (\ref{S}) and (\ref{ZT}) become
\beq S(T\to \infty)= \frac{\epsilon_0-\mu + U/2}{eT} \equiv \frac{V_g}{T}\label{Sinf}\eeq
and, further using (\ref{Bld0}), 
\beq ZT(T\to \infty) = \frac{4V^2_g}{U^2 +2D\Gamma/\pi}.\label{ZTinf}\eeq
Here, we have introduced the \emph{gate voltage} 
\beq V_g \equiv (\epsilon_0-\mu + U/2)/e.\label{gV}\eeq 

Taking $D=0$ or $\Gamma=0$, (\ref{Sinf}) and (\ref{ZTinf}) reproduce the analytic high-temperature limiting values found in Ref.~\onlinecite{Murphy06} [specifically, their Eqs.~(19) and (21)] using the sequential tunnelling approximation~\cite{Beenakker92,Koch04} for $\Gamma\simeq0$.  From (\ref{ZTinf}), one sees that the diverging figure of merit of Mahan and Sofo (valid for all temepratures) is realized in the noninteracting $U\ll \epsilon_0$ and \emph{un-broadened} limit $\B,\Gamma\ll \epsilon_0$.   The high-$T$ limiting form (\ref{Sinf}) for the Seebeck coefficient has the same form as the ``Mott-Heikes'' approximation $S(T\to \infty) = [\mu(T=0)-\mu]/eT$---well-known in studies of bulk narrow-band systems~\cite{Peterson10,Chaikin76}---with the identification $\mu(T=0)=\epsilon_0 + U/2$, which is indeed the zero-temperature chemical potential at half-filling, $n_d=1$.

Although these high-temperature results are valid in a temperature regime well-outside that found in typical systems, they can serve as useful checks of numerical calculations.  Any conserving calculation~\cite{Baym61} results in a transmission function (single-particle Green's function) that satisfies the sum rules and hence, (\ref{Sinf}) and (\ref{ZTinf}).  As we have already noted, the sequential tunnelling approximation used in Ref.~\onlinecite{Murphy06} satisfies these asymptotic results at high temperatures.  Our results also give analytic expressions for the high-temperature asymptotes found numerically in Ref.~\onlinecite{Costi10} using numerical renormalization group~\cite{Bulla08}.

\section{Thermoelectric performance in the strongly correlated regime $U\gg T$.}
\label{midTsec}
In the strong-correlation, narrow-level regime $U\gg T \gtrsim \Gamma$, (\ref{M0t})-(\ref{M2t}) give
\beq S(U\gg T\gtrsim \Gamma) = \frac{1}{eT}\left[\epsilon_0 + \frac{\langle {\hat{\cal{V}}}^{\dagger}\rangle}{2-n_d}-\mu + {\cal{O}}\left(\tfrac{V^2}{\epsilon_0+U}\right)\right]\label{Sc}\eeq
for the thermopower and
\begin{align}
ZT = &\frac{(\epsilon_0-\mu)^2(2-n_d)^2 + 2(\epsilon_0-\mu)\langle {\hat{\cal{V}}}^{\dagger}\rangle (2-n_d)+\langle {\hat{\cal{V}}}^{\dagger}\rangle^2}{(2-n_d)^2\sum_{\bk\nu}|V_{\bk\nu}|^2+(2-n_d)[\langle {\hat{\cal{V}}}^{\dagger}\hat{\cal{V}}\rangle -\langle\tilde{\cal{V}}^{\dagger}\xi\rangle] - \langle {\hat{\cal{V}}}^{\dagger}\rangle^2}\nonumber\\&+ {\cal{O}}\left(\tfrac{V}{\epsilon_0+U}\right),\label{ZTc}\end{align} 
for the figure of merit, where we have defined ($\xi \equiv \epsilon-\mu$)
\beq \langle\tilde{\cal{V}}^{\dagger}\xi\rangle \equiv \sum_{\bk\sigma\nu}V^{*}_{\bk\nu\sigma}(\epsilon_{\bk\nu}-\mu)\langle\hat{d}^{\dagger}_{\sigma}\hat{c}_{\bk\nu\sigma}\rangle.\label{Ald}\eeq
Note that we are using the bare electron operators in the above and $\langle\hat{\cal{V}}\rangle$ is defined using (\ref{v}).  Although (\ref{Sc}) and (\ref{ZTc}) are asymptotically exact in the limit where there is a large separation between energy scales, $\Gamma\ll T\ll U$ (also the limit where thermoelectric performance is optimized), as noted earlier, we expect them to provide a reasonable approximation when $\Gamma \lesssim T$.

\begin{center}
\begin{figure}
\includegraphics[width=0.8\linewidth]{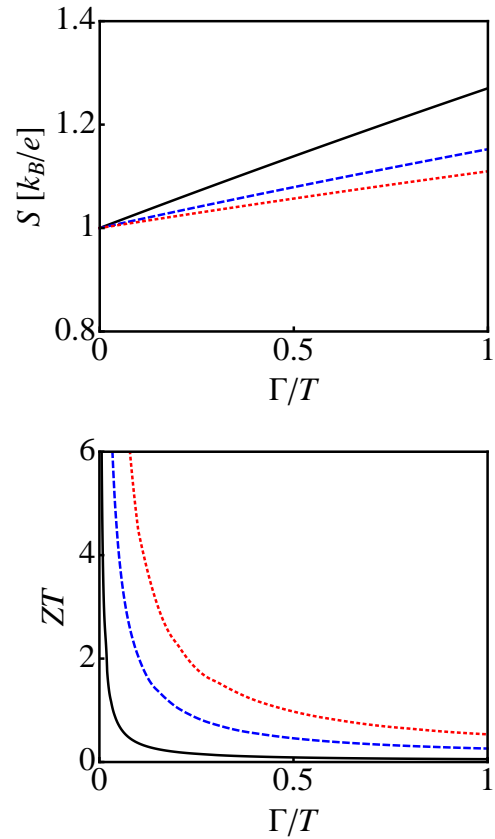}
\caption{Dependence of the Seebeck coefficient (top) and the electron contribution to the figure of merit (bottom) on the broadening $\Gamma$ for different lead bandwidths: $D/T=100$ (black solid line), $D/T=20$ (blue dashed line), and $D/T=10$ (red dotted line).  Curves are shown for $U=100T$ and $\epsilon_0-\mu = T$ although the behaviour shown here is not sensitive to the precise choice of values.}\label{HFfig}
\end{figure}
\end{center}

The thermopower (\ref{Sc}) in the strongly-correlated regime and intermediate temperatures ($T\gtrsim \Gamma$) again has the Mott--Heikes form, but with a renormalized low-energy measure $\epsilon_0 +  \langle {\hat{\cal{V}}}^{\dagger}\rangle/(2-n_d)$ of the single-particle energy that does not include the Coulomb energy scale.    In fact, apart from the renormalization factor $2-n_d$ accounting for the absence of doubly-occupied states, this is precisely the single-particle energy of a noninteracting electron in an impurity coupled to leads by (\ref{V}).  Consistent with the findings of Ref.~\onlinecite{Murphy06}, by eliminating the possibility of having a doubly occupied impurity, extremely strong correlations have effectively reduced the transport problem to that of a single-level.  As a result, transport is effectively that of noninteracting electrons, albeit with a reduced Hilbert space ($n_d\neq 2$)

As with noninteracting electrons, thermoelectric performance in this regime is limited by the broadening due to the coupling between the impurity and the lead.  Specifically, when the lead-impurity coupling is larger than the detuning from the Fermi level, $\langle{\hat{\cal{V}}}\rangle \gtrsim \epsilon_0-\mu$, the figure of merit (\ref{ZTc}) is generically of order unity.  It is only when the coupling becomes much smaller than the detuning, $\langle{\hat{\cal{V}}}\rangle \ll \epsilon_0-\mu$, that one obtains the diverging electronic contribution to the figure of merit found in Ref.~\onlinecite{Murphy06}.  Outside this limit, even though the strong Coulomb repulsion has reduced the problem to an effectively single-level one, the variance in the single-particle energy (roughly, the greater of $\sim \sum_{\bk\nu}|V_{\bk\nu}|^2$ and $\langle {\hat{\cal{V}}}^{\dagger}\hat{\cal{V}}\rangle - \langle {\hat{\cal{V}}}^{\dagger}\rangle^2$) prevents the system from attaining optimal thermoelectric performance.  

All terms in the denominator of (\ref{ZTc}) vanish in the $\Gamma\to 0$ \emph{or} $\B\to 0$ limit, meaning that---barring a nonanalytic dependence on either of these variables---the figure of merit will exhibit the scaling  shown in (\ref{ZTdiv}) for small $\Gamma,\B$.  To check this, in Fig.~\ref{HFfig} we show (\ref{Sc}) and (\ref{ZTc}) as functions of $\Gamma$ using Hartree--Fock (HF) theory and (\ref{Bld0}) for a range of bandwidths.  Details of the HF calculations are given in Appendix~\ref{HFsec}.  Although HF is not a quantitatively reliable theory when interactions are strong, we emphasize that we are only applying HF theory to evaluate the asymptotically exact (conserving) results (\ref{Sc}) and (\ref{ZTc}), and we expect it to give qualitatively reliable results.  We are \emph{not} presenting the results of a (non-conserving) HF theory of transport, the results of which would be completely unreliable.  While the thermopower is generically of order unity (in units of $k_B/e$) for $\epsilon_0-\mu = T$, the figure of merit is singular at small broadenings and is well-described by the scaling in (\ref{ZTdiv}).  For comparison with this plot,  in molecular systems, $\Gamma\sim(0.01\to0.5)$eV~\cite{Quek07,Paulsson03} or, $\Gamma = (0.4\to19)T_{\mathrm{room}}$, with $T_{\mathrm{room}} = 300K$.  For leads constructed of a good metal such as gold, $D\sim 1\mathrm{eV} \sim 40 T_{\mathrm{room}}$.  

Fig.~\ref{HFfig} emphasizes that either $\Gamma$ or the lead bandwidth $\B$ need to be much smaller than temperature in order to reproduce the substantial figure of merit found in Ref.~\onlinecite{Murphy06} for $U\to \infty$.   Reinforcing the fact that the thermoelectric transport properties in the very strongly correlated regime $U\gg T$ is equivalent to that of a renormalized non-interacting system, we note that Fig.~\ref{HFfig} closely resembles the $U=0$ Fig.~1 in  Ref.~\onlinecite{Murphy06}.

\section{Discussion}
\label{discussionsec}

Much of the interest in studying the thermoelectric performance of quantum dots and molecules originates in the likely small phonon contribution to the thermal conductance in these systems and also the fact that they can be well-approximated as single-level systems.  For a single-level noninteracting impurity system, the figure of merit diverges in the limit that the extrinsic broadening arising from the coupling to the leads is much smaller than the temperature.  Turning on a Coulomb repulsion, the appearance of a satellite Hubbard peak leads to an increase in the variance of the single-particle energy level and hence, a diminished figure of merit.  

Somewhat counterintuitively, when the Coulomb repulsion is \emph{very} strong,  however, the new satellite Hubbard level becomes unimportant, as it is too high an energy scale to be relevant for transport.  We have used asymptotically exact sum rule expressions in this limit to show that the thermoelectric transport coefficients assume essentially the same form as for a noninteracting system, slightly renormalized to account for the absence of doubly occupied states.  The thermopower (\ref{Sc}) in particular assumes the well-known Mott--Heikes form, but with an effective noninteracting chemical potential.  As with a noninteracting system, the transport properties in this strong interaction regime are limited only by the broadening due to the coupling between the leads and the impurity.

This bodes well for using small impurity-type systems such as quantum dots and molecules to achieve substantial thermoelectric efficiencies.  Even though the bare (un-screened) Coulomb energy scale is very large ($U\gg T$) in these systems, it is precisely its large size that means it does not lead to a diminishment of transport, as would happen if it was of moderate strength, say $U\sim T$.  At the same time, the highly singular dependence of the figure of merit on the broadening and also the lead bandwidth [c.f. (\ref{ZTdiv})] means that it will be essential to have a small broadening arising from lead-impurity coupling or to tailor the bandwidth properties of the leads in order to realize the potential of impurity systems.

\acknowledgements
This work was funded by the Natural Sciences and Engineering
Research Council of Canada Discovery Grant and the Canada Research Chair Program.

\appendix

\section{Spectral properties of the single-particle Green's function}
\label{spectralsec}

In this Appendix, we give an example of the derivation of the general result (\ref{Minfexp}) for $n=1$:
\begin{align} &M^{T}_1\equiv \int^{\infty}_{-\infty}dE E A(E)\nonumber\\
&=\sum_{\sigma}\sum_{a,b}(P_a+P_b)\langle a|\hat{d}_{\sigma}|b\rangle\langle b| \hat{d}^{\dagger}_{\sigma} |a\rangle (E_{b}-E_a)\nonumber\\
&=\sum_{\sigma}\sum_{a,b}P_a\left[\langle a|\hat{d}_{\sigma}\hat{H} |b\rangle \langle b |\hat{d}^{\dagger}_{\sigma} |a\rangle -\langle a|\hat{H}\hat{d}_{\sigma} |b\rangle \langle b |\hat{d}^{\dagger}_{\sigma} |a\rangle \right]\nonumber\\
&+\sum_{\sigma}\sum_{a,b} P_b\left[\langle b |\dsdj |a\rangle\langle a|\ds\hat{H} |b\rangle -\langle b |\dsdj |a\rangle\langle a|\hat{H}\ds |b\rangle \right]\nonumber\\
&=\sum_{\sigma}\sum_{a}P_a\left[\langle a|\ds\hat{H} \dsd |a\rangle -\langle a|\hat{H}\ds\dsdj |a\rangle \right]\nonumber\\
&+\sum_{\sigma}\sum_{b} P_b\left[\langle b |\dsdj \ds\hat{H} |b\rangle -\langle b |\dsdj \hat{H}\ds |b\rangle \right]\nonumber\\
&=\sum_{\sigma}\left[\langle \ds\hat{H} \dsdj \rangle - \langle \hat{H}\ds \dsdj\rangle + \langle \dsdj \ds \hat{H}\rangle - \langle \dsdj\hat{H}\ds\rangle\right]\nonumber\\ &=\sum_{\sigma}\langle\{ \dsdj, [\ds,\hat{H}]\}\rangle.
\end{align}

\section{Hartree--Fock theory for the AIM model}
\label{HFsec}

In this Appendix, we describe how to calculate the impurity occupation $n_d$, lead-impurity coherence 
$\langle \hat{\cal{V}}\rangle = \langle \hat{\cal{V}}^{\dagger}\rangle$, and $\langle\tilde{\cal{V}}^{\dagger}\xi\rangle$, defined in (\ref{Ald}) using Hartree--Fock (HF) theory.  Within this approximation, $\langle \hat{\cal{V}}^{\dagger}\hat{\cal{V}}\rangle = \langle \hat{\cal{V}}\rangle^2$ and hence, these three quantities are sufficient to evaluate the figure of merit (\ref{ZTc}).   

Momentarily generalizing the AIM hamiltonian (\ref{AIM}) to allow for different chemical potentials and couplings for each spin, $\mu\to \mu_{\sigma}$ and $V_{\bk\nu}\to V_{\bk\nu\sigma}$, one has
\beq n_{\sigma} \equiv \langle\hat{n}_{\sigma}\rangle = -\left(\frac{\partial \Omega}{\partial\mu_{\sigma}}\right)_{T}\label{nsigformal}\eeq
and
\beq \langle \hat{d}_{\sigma}\hat{c}^{\dagger}_{\bk\nu\sigma}\rangle=-\langle \hat{c}^{\dagger}_{\bk\nu\sigma}\hat{d}_{\sigma}\rangle = -\left(\frac{\partial \Omega}{\partial V_{\bk\nu\sigma}}\right)_{T},\label{coherenceformal}\eeq
where $\Omega$ is the free energy.  At the end of our calculation, we set $\mu_{\uparrow}=\mu_{\downarrow}=\mu$ and likewise with $V_{\bk\nu\sigma}$.  (\ref{nsigformal}) is a standard thermodynamic identity, while (\ref{coherenceformal}) follows from an application of the Hellmann--Feynman formula $(\partial \Omega /\partial \lambda)_{T,N} = \langle \partial\hat{H}/\partial\lambda\rangle$ with $\lambda=V_{\bk\nu\sigma}$ to (\ref{AIM}).  

Since it is an effective single-particle theory, the HF free energy is given by the trace of the logarithm of the HF Green's function $G_{\sigma}(i\omega_n)$ for spin $\sigma$ impurity electrons: $\Omega = -\beta^{-1}\sum_{n\sigma}e^{i\omega_n0^+}\ln [-G^{-1}_{\sigma}(i\omega_n)]$, yielding
\beq \Omega =-\frac{1}{\beta}\sum_{n\sigma}e^{i\omega_n0^+}\ln\Big[-i\omega_n + \epsilon_0-\mu_{\sigma}  + Un_{\bar{\sigma}}+\sum_{\bk\nu}\tfrac{|V_{\bk\nu\sigma}|^2}{i\omega_n-\xi_{\bk \nu \sigma}}\Big].\label{HF1}\eeq
Here $\omega_n = (2n+1)/\beta$ are Fermi Matsubara frequencies, with integer $n$, and $\xi_{\bk\nu\sigma}\equiv \epsilon_{\bk\nu}-\mu_{\sigma}$. The $\exp(i\omega_n 0^+)$ factor with positive infinitesimal $0^+$ ensures convergence.  
Making the same assumptions (constant lead DOS, momentum-independent coupling $V_{\bk\nu\sigma}$) that we used to arrive at (\ref{GammaAIM2}), (\ref{HF1}) reduces to
\beq \Omega \!=\!-\frac{1}{\beta}\sum_{n\sigma}e^{i\omega_n0^+}\!\ln\!\Big[\!-i\omega_n\! + \epsilon_0-\mu_{\sigma}+ Un_{\bar{\sigma}}- \left(\!\tfrac{\Gamma}{4\pi}\!\right)\!\ln\! \left(\!\tfrac{i\omega_n - \B}{i\omega_n+\B}\!\right)\!\Big].\eeq
We will not make the usual approximation in the AIM literature of taking $D\to \infty$, in which case 
$\ln [(i\omega_n-\B)/(i\omega_n+\B)]$ reduces to $\pi\mathrm{sgn} (\omega_n)$ (see e.g., Sec. 5.2 in Ref.~\onlinecite{Hewsonbook}).  
  
Applying (\ref{nsigformal}) and (\ref{coherenceformal}) to (\ref{HF1}) [for the coherence (\ref{coherenceformal}) we need to take the derivative of the free energy with respect to $V_{\bk\nu\sigma}$ \emph{before} assuming the coupling to be constant], one arrives at the results
\beq n_{\sigma} =\frac{1}{\beta}\sum_n \frac{ e^{i\omega_n0^+}}{i\omega_n - \epsilon_0+\mu_{\sigma} - Un_{\bar{\sigma}} + \left(\tfrac{\Gamma}{4\pi}\right)\ln \left(\tfrac{i\omega_n - \B}{i\omega_n+\B}\right)}\label{HFn}\eeq
and
\begin{align} &\langle \hat{d}_{\sigma}\hat{c}^{\dagger}_{\bk\nu\sigma}\rangle =-\frac{1}{\beta}\sum_n \frac{e^{i\omega_n0^+}}{\left(i\omega_n-\xi_{\bk\sigma}\right)} \nonumber\\
&\times \frac{V^{*}_{\bk\nu\sigma}}{\left[i\omega_n - \epsilon_0+\mu_{\sigma} - Un_{\bar{\sigma}} + \left(\tfrac{\Gamma}{4\pi}\right)\ln \left(\tfrac{i\omega_n - \B}{i\omega_n+\B}\right)\right]}\end{align}
Using this last result to evaluate $\langle\hat{\cal{V}}\rangle$ and (\ref{Ald}) and replacing the momentum integration with one over energy [assuming again, a constant lead DOS as in (\ref{bandwidthlimapprox})] gives
\beq \langle\hat{\cal{V}}\rangle= \frac{1}{\beta}\sum_n\frac{ e^{i\omega_n0^+}\left(\tfrac{\Gamma}{2\pi}\right)\ln \left(\tfrac{i\omega_n - \B}{i\omega_n+\B}\right)}{i\omega_n - \epsilon_0+\mu_{\sigma} - Un_{\bar{\sigma}} +  \left(\!\tfrac{\Gamma}{4\pi}\!\right)\!\ln\! \left(\!\tfrac{i\omega_n - \B}{i\omega_n+\B}\!\right)}\label{HFV}\eeq
and
\beq \langle\tilde{\cal{V}}^{\dagger}\xi\rangle = \frac{1}{\beta}\sum_n\frac{ e^{i\omega_n0^+}\!\left(\tfrac{\Gamma}{2\pi}\right)\!\left[2\B + i\omega_n\ln \left(\tfrac{i\omega_n - \B}{i\omega_n+\B}\right)\!\right]}{i\omega_n - \epsilon_0+\mu_{\sigma} - Un_{\bar{\sigma}} +  \left(\!\tfrac{\Gamma}{4\pi}\!\right)\!\ln\! \left(\!\tfrac{i\omega_n - \B}{i\omega_n+\B}\!\right)},\label{HFA}\eeq

To obtain the results shown in Fig.~\ref{HFfig}, we solve (\ref{HFn}) self-consistently for $n_{\sigma}$, evaluating the Matsubara frequency sum numerically by summing between $n=-10^4\to 10^4$.  To avoid dealing with the convergence factor, we use
\beq n_{\sigma} =\frac{1}{2}+\frac{1}{\beta}\sum_n \frac{1}{i\omega_n - \epsilon_0+\mu_{\sigma} - Un_{\bar{\sigma}} + \left(\tfrac{\Gamma}{4\pi}\right)\ln \left(\tfrac{i\omega_n - \B}{i\omega_n+\B}\right)}.\label{HFnb}\eeq
This can be derived from (\ref{HFn}) by re-writing the frequency sum as an integral over the complex plane and manipulating the contour of integration; see e.g. Appendix A in Ref.~\onlinecite{Ohashi03}.  In contrast to (\ref{HFn}), for which the convergence factor is needed to deal with a summand that vanishes as $(i\omega_n)^{-1}$ at large frequencies, the summands in both (\ref{HFV}) and (\ref{HFA}) vanish as $(i\omega_n)^{-2}$, and we can simply drop the convergence factor.  The solution of (\ref{HFnb}) is used in these and we again sum over the Matsubara frequencies, from $n=-10^4\to 10^4$, to obtain $\langle \hat{\cal{V}}\rangle$ and $\langle\tilde{\cal{V}}^{\dagger}\xi\rangle$.


\begin{thebibliography}{99} 
\bibitem{Mahan96} G.~D.~Mahan and J.~O.~Sofo, Proc. Nat. Acad. Sci. \textbf{93}, 7436 (1996).
\bibitem{Scheibner05} R.~Scheibner, H.~Buhmann, D.~Reuter, M.~N.~Kiselev, and L.~W.~Molenkamp, \href{http://dx.doi.org/10.1103/PhysRevLett.95.176602}{Phys. Rev. Lett. \textbf{95}, 176602 (2005).}
\bibitem{Costi10} T. A. Costi and V. Zlati\'c, \href{http://dx.doi.org/10.1103/PhysRevB.81.235127}{Phys. Rev. B \textbf{81}, 235127 (2010).}
\bibitem{Haupt13} F.~Haupt, M.~Leijnse, H.~L.~Calvo, L.~Classen, J.~Splettstoesser, and M.~R.~Wegewijs, 
\href{http://dx.doi.org/10.1002/pssb.201349219}{Phys. Status Solidi B \textbf{250}, 2315 (2329).}
\bibitem{Sothmann15} B.~Sothmann, R.~S\'anchez, and A.~N.~Jordan, \href{http://dx.doi.org/10.1088/0957-4484/26/3/032001}{Nanotechnology \textbf{26}, 032001 (2015).}
\bibitem{Paulsson03} M.~Paulsson and S.~Datta, \href{http://dx.doi.org/10.1103/PhysRevB.67.241403}{Phys. Rev. B \textbf{67}, 241403(R) (2003).}
\bibitem{Reddy07} P.~Reddy, S.-Y.~Jang, R.~A.~Segalman, and A.~Majumdar, \href{http://dx.doi.org/10.1126/science.1137149}{Science \textbf{315}, 1568 (2007).}
\bibitem{Malen10} J.~A.~Malen, S.~K.~Yee, A.~Majumdar, and R.~A.~Segalman, \href{http://dx.doi.org/10.1016/j.cplett.2010.03.028}{Chem. Phys. Lett. \textbf{491}, 109 (2010).}
\bibitem{Andergassen11}  S.~Andergassen, T.~A. Costi, and V.~Zlati\'c, \href{http://dx.doi.org/10.1103/PhysRevB.84.241107}{Phys. Rev. B \textbf{84}, 241107(R) (2011).}
\bibitem{Dubi11} Y.~Dubi and M.~Di Ventra, \href{http://dx.doi.org/10.1103/RevModPhys.83.131}{Rev. Mod. Phys. \textbf{83}, 131 (2011).}  
\bibitem{Yee11} S.~K.~Yee, J.~A.~Malen, A.~Majumdar, and R.~A.~Segalman, \href{http://dx.doi.org/10.1021/nl2014839}{Nano Lett.  \textbf{11}, 4089 (2011).}
\bibitem{Lee13} W.~Lee, K.~Kim, W.~Jeong, L.~A.~Zotti, F.~Pauly, J.~C.~Cuevas, and P.~Reddy, \href{http://dx.doi.org/10.1038/nature12183}{Nature \textbf{498}, 209 (2014).} 
\bibitem{Kim14} Y.~Kim, W.~Jeong, K.~Kim, W.~Lee, and P.~Reddy, \href{http://dx.doi.org/10.1038/NNANO.2014.209}{Nature Nanotech. \textbf{9}, 881 (2014).}
\bibitem{Stadler11} R.~Stadler and T.~Markussen, \href{http://dx.doi.org/10.1063/1.3653790}{J. Chem. Phys. \textbf{135}, 154109 (2011).}
\bibitem{Burkle15} M.~B\"urkle, T.~J.~Hellmuth, F.~Pauly, and Y.~Asai, \href{http://dx.doi.org/10.1103/PhysRevB.91.165419}{Phys. Rev. B \textbf{91}, 165419 (2015).}
\bibitem{Costi94} T.~A.~Costi, A.~C.~Hewson, and V.~Zlati\'c, \href{http://dx.doi.org/10.1088/0953-8984/6/13/013}{J. Phys. Condens. Matter \textbf{6}, 2519 (1994).}
\bibitem{Liu10} J.~Liu, Q.-f.~Sun, and X.~C.~Xie, \href{http://dx.doi.org/10.1103/PhysRevB.81.245323}{Phys. Rev. B \textbf{81}, 245323 (2010).}
\bibitem{Ren12} J.~Ren, J.-X.~Zhu, J.~E.~Gubernatis, C.~Wang, and B.~Li, \href{http://dx.doi.org/10.1103/PhysRevB.85.155443}{Phys. Rev. B \textbf{85}, 155443 (2012).}
\bibitem{Beenakker92} C.~W.~J.~Beenakker and A.~A.~M.~Staring, \href{http://dx.doi.org/10.1103/PhysRevB.46.9667}{Phys. Rev. B \textbf{46}, 9667 (1992).}  
\bibitem{Koch04} J.~Koch, F.~von Oppen, Y.~Oreg, and E.~Sela, \href{http://dx.doi.org/10.1103/PhysRevB.70.195107}{Phys. Rev. B \textbf{70}, 195107 (2004).} 
\bibitem{Murphy06} P.~Murphy, S.~Mukerjee, and J.~Moore, \href{http://dx.doi.org/10.1103/PhysRevB.78.161406}{Phys. Rev. B \textbf{78}, 161406(R), (2008).}
\bibitem{Beni74} G.~Beni, \href{http://dx.doi.org/10.1103/PhysRevB.10.2186}{Phys. Rev. B \textbf{10}, 2186 (1974).}
\bibitem{Chaikin76} P.~M.~Chaikin and G.~Beni, \href{http://dx.doi.org/10.1103/PhysRevB.13.647}{Phys. Rev. B \textbf{13}, 647 (1976).}
\bibitem{Mukerjee07} S.~Mukerjee and J.~E.~Moore, \href{http://dx.doi.org/10.1063/1.2712775}{App. Phys. Lett. \textbf{90}, 112107 (2007).}
\bibitem{Baym61} G.~Baym and L.~P.~Kadanoff, \href{http://dx.doi.org/10.1103/PhysRev.124.287}{Phys. Rev. \textbf{124}, 287 (1961).}
\bibitem{Benenti13} G.~Benenti, G.~Casati, T.~Prosen, and K.~Saito, \href{http://arxiv.org/abs/1311.4430}{arXiv:1311.4430.}
\bibitem{MW} Y.~Meir and N.~S.~Wingreen, \href{http://dx.doi.org/10.1103/PhysRevLett.68.2512}{Phys. Rev. Lett. \textbf{68}, 2512 (1992).}
\bibitem{Kim02} T.-S.~Kim and S.~Hershfield, \href{http://dx.doi.org/10.1103/PhysRevLett.88.136601}{Phys. Rev. Lett. \textbf{88}, 136601 (2002).}
\bibitem{Quek07} S.~Y.~Quek, L.~Venkataraman, H.~J.~Choi, S.~G.~Louie, M.~S.~Hybertsen, and J.~B.~Neaton, \href{http://dx.doi.org/10.1021/nl072058i}{Nano Lett. \textbf{7}, 3477 (2007).}
\bibitem{White91} S.~R.~White, \href{http://dx.doi.org/10.1103/PhysRevB.44.4670}{Phys. Rev. B \textbf{44}, 4670 (1991).}
\bibitem{Harris67} A.~B.~Harris and R.~V.~Lange, \href{http://dx.doi.org/10.1103/PhysRev.157.295}{Phys. Rev. \textbf{157}, 295 (1967).}
\bibitem{kappacomment} This bound actually furnishes a much stronger bound for the thermal conductance $\kappa_0$ at zero chemical potential (voltage) difference, $\mu_L=\mu_R$, since $ \kappa_0 = (\Gamma/8\hbar  T^2)M_2$.  
\bibitem{Taylor15} E.~Taylor and D.~Segal, \href{http://dx.doi.org/10.1103/PhysRevLett.114.220401}{Phys. Rev. Lett. \textbf{114}, 220401 (2015).}
\bibitem{comment} By ``negligible'' here, we mean $E^2A(E)\ll T^2$ for $\Gamma\ll E\ll U$, since this quantity enters the $n=2$ sum rules.  Since the high-frequency tail $\lim_{E\gg U}A(E)\propto E^{-2}$ recovers the noninteracting Lorentzian asymptote, one also requires the lead bandwidth $\B$ to be $\lesssim U$ in order to satisfy this constraint.  
\bibitem{Hewsonbook} A.~C.~Hewson, \textit{The Kondo Problem To Heavy Fermions, Cambridge Studies in Magnetism}, (Cambridge University Press, Cambridge, England, 1997).
\bibitem{OPE} For examples of sum-rule calculations involving non-standard weight functions using the OPE, see: E.~C.~Poggio, H.~R.~Quinn, and S.~Weinberg, \href{http://dx.doi.org/10.1103/PhysRevD.13.1958}{Phys. Rev. D \textbf{13}, 1958 (1976)} and E.~Braaten, D.~Kang, and L.~Platter, \href{http://dx.doi.org/10.1103/PhysRevLett.104.223004}{Phys. Rev. Lett. \textbf{104}, 223004 (2010).}  
\bibitem{Eskes94} H.~Eskes , A.~M.~Ole\'s, M.~B.~J.~Meinders, and W.~Stephan, \href{http://dx.doi.org/10.1103/PhysRevB.50.17980}{Phys. Rev. B \textbf{50}, 17980 (1994).}
\bibitem{Randeria05} M.~Randeria, R.~Sensarma, N.~Trivedi, and F.-C.~Zhang, \href{http://dx.doi.org/10.1103/PhysRevLett.95.137001}{Phys. Rev. Lett. \textbf{95}, 137001 (2005).}
\bibitem{Schrieffer66} J.~R.~Schrieffer and P.~A.~Wolff, \href{http://dx.doi.org/10.1103/PhysRev.149.491}{Phys. Rev. \textbf{149}, 491 (1966).} 
\bibitem{MacDonald88} A.~H.~MacDonald, S.~M.~Girvin, and D.~Yoshioka, \href{http://dx.doi.org/10.1103/PhysRevB.37.9753}{Phys. Rev. B \textbf{37}, 9753 (1988).}
\bibitem{SWcomment} This result only differs from the Schrieffer--Wolff transformation in the form of the denominator which for us, does not include the kinetic energy of the bath electrons.  This difference---unimportant for the final result---stems from our having included the corresponding term in the hamiltonian with the other hopping terms, while Schrieffer and Wolff keep them in $\hat{H}_0$.  
\bibitem{Peterson10} M.~R.~Peterson and B.~S.~Shastry, \href{http://dx.doi.org/10.1103/PhysRevB.82.195105}{Phys. Rev. B \textbf{82}, 195105 (2010).} 
\bibitem{Bulla08} R.~Bulla, T.~A.~Costi, and T.~Pruschke, \href{http://dx.doi.org/10.1103/RevModPhys.80.395}{Rev. Mod. Phys. \textbf{80}, 395 (2008).}
\bibitem{Ohashi03} Y.~Ohashi and A.~Griffin, \href{http://dx.doi.org/10.1103/PhysRevA.67.033603}{Phys. Rev. A \textbf{67}, 033603 (2003).}
\end{thebibliography}
\end{document}